\documentstyle[pra,aps]{revtex}
\begin{document}
\draft

\title{INTERPRETATION    OF   THE  EVOLUTION  PARAMETER  OF    THE    FEYNMAN
PARAMETRIZATION OF THE DIRAC EQUATION}

\author{Juan P.  APARICIO,$^1$ Fabi\'an H.  GAIOLI,$^{1,2,}$\footnote{e-mail:
gaioli@iafe.uba.ar,  \  fax:    54  1  786  8114}  and  Edgardo  T.    GARCIA
ALVAREZ$^{1,2,}$\footnote{e-mail:  galvarez@dfuba.df.uba.ar, \ fax:  54 1 786
8114}}

\address{$^1$Departamento  de  F\'{\i}sica,  Facultad  de Ciencias Exactas  y
Naturales, \\ Universidad de Buenos Aires, 1428 Buenos Aires, Argentina}

\address{$^2$Instituto  de Astronom\'\i a y F\'\i sica del Espacio,  \\  C.C.
67, Suc.  28, 1428 Buenos Aires, Argentina}

\date{\today}

\maketitle

\begin{abstract}

The Feynman  parametrization  of the Dirac equation is considered in order to
obtain an indefinite  mass formulation of relativistic quantum mechanics.  It
is shown that the  parameter  that  labels  the  evolution  is related to the
proper time.  The St\"uckelberg  interpretation  of  antiparticles  naturally
arises from the formalism.

\end{abstract}
\vskip 2cm

\twocolumn
\narrowtext

Relativistic quantum mechanics (RQM) parametrized by  a  ``proper  time'' has
been a powerful device used long ago.    The  leading  idea  of a proper time
formalism  rests  on  considering  states  that evolve with  a  Schr\"odinger
equation  with  a  ``scalar''  Hamiltonian  which plays the role  of  a  mass
operator.   The framework, which provides indefinite mass states as  well  as
the  St\"uckelberg  \cite{Stu}  interpretation    for  antiparticles,  allows
avoiding the well-known difficulties of  RQM  \cite{Fan}.    In  other words,
admitting  particles moving backward in time,  we  may  keep  a  one-particle
formalism  without appealing to the standard solution  given  by  the  second
quantization scheme \cite{Dav}.

The origin of this subject goes back to the earlier works of Dirac \cite{Di},
Fock  \cite{Fo},  St\"uckelberg  \cite{Stu},  Feynman  \cite{f50,Fey},  Nambu
\cite{Na},  and  Schwinger  \cite{Schw}.   More recently,  in  this  line  of
research we can mention the relativistic dynamics (RD)  by  Horwitz  {\it  et
al.}  \cite{H},  the  four  space formulation (FSF) by Fanchi  {\it  et  al.}
\cite{F}, and the works of the French school (Vigier {\it  et al.} \cite{V}),
mainly  developed  in  the  spin  $0$  case \cite{exc}, which have in  common
squared mass operators as Hamiltonians.

RD and FSF approaches use a Fock-like parametrization given by

\begin{equation}
-i \frac{\partial}{\partial  \tau} \Phi = \frac{p^{\mu} p_{\mu}}{2M} \Phi \ \
\ \ \ (c=\hbar=1)
\label{faho}
\end{equation}
in the free case.  This parametrization is very interesting because it can be
seen  as  a  representation  of a five dimensional Galilei group  \cite{Rom},
which introduces a sort of ``super-mass'' $M$ (with units of mass)  as  a new
label for  characterizing  the  indefinite  mass system mentioned above.  The
French    school    uses    the    St\"uckelberg-Schwinger    \cite{Stu,Schw}
parametrization \cite{DV}, which can  be  obtained  from Eq.  (\ref{faho}) by
rescaling the dimension of the  evolution  parameter  $\tau$  and  taking $M$
equal to $1/2$.

It   is  well  known  that,  taking  the  ``on-shell''  condition\footnote{By
``on-shell'' condition  we  mean  a  reinterpretation of the usual mass-shell
constraint as a  result  of the specific initial conditions.  In other words,
$\langle p^{\mu} p_{\mu} \rangle_c$  is a classical constant of motion [where
the subscript $c$ denotes classical  ($\hbar\rightarrow  0$)  mean value, see
Appendix], which acts as a square  mass  variable  that  can  be  fixed  to a
particular value $m_0^2$, being $m_0$ the ordinary  mass  of the particle, by
choosing the initial conditions [$\langle x^{\mu}(0) \rangle_c$ and  $\langle
p_{\mu}(0)  \rangle_c$].} in the classical limit, one can recover  the  usual
relativistic  mechanics   from  these  parametrizations.    However,  in  the
classical limit the  evolution parameter $\tau$ is related to the proper time
$s$ by means of $ds^2 = (m_0^2/M^2) d\tau^2$.  Therefore, $\tau$ is not equal
to $s$ unless one identifies $\langle p^{\mu} p_{\mu} \rangle_c = m_0^2$ with
$M^2$ \cite{H,F}, where $\langle \ \rangle_c$ stands for the mean value after
taking the classical limit (see Appendix).

The main problem of the different proposals of a parametrized RQM lies on the
``off-shell''  interpretation  of  the evolution parameter $\tau$ \cite{fa3}.
The  different names proposed (within or without interpretation) reflect  the
controversy about this subject.  In the past, this parameter was treated only
by analogy as proper time, but it is not its accurate  sense since it must be
interpreted as a Newtonian time \cite{hfp}.

The aim of  this letter is to show that, by using a different parametrization
from the given in  Eq.   (\ref{faho}), an ``off-shell'' interpretation of the
corresponding evolution parameter as the  proper time can be given.  In fact,
by  considering  a  first  order  mass    operator    corresponding   to  the
parametrization  of  the  Dirac  equation  originally  proposed   by  Feynman
\cite{Fey} we show that, in the classical limit,  the  ``evolution  time'' of
this  parametrization  is reduced to the proper time of  an  indefinite  mass
system.  As we will show in Ref.  \cite{deS}, unlike the interpretation given
in  Ref.    \cite{Rom}  to  the  parametrization  (\ref{faho}),  the  Feynman
parametrization can be  looked as a null ``super-mass'' representation of the
de Sitter group.  It immediately leads to the identification of the evolution
parameter with the proper time $s$ since in this case the arc element $dS$ of
the five-dimensional manifold associated to the  de  Sitter  group  vanishes,
i.e., $dS^2 = ds^2 -dx^{\mu}dx_{\mu} = 0$  \cite{deS}.   For these reasons we
call $s$ the evolution parameter of the Feynman  parametrization  since it is
directly  related  to  the  classical proper time $s$, unlike  the  parameter
$\tau$.   However, it is important to remark that, although  proper  time  is
commonly  considered  as an on-shell concept since it is associated with  the
integral  of  the  arc  element along the world line of the standard  massive
particles,  we  can  extend  this  concept  using  the  same  definition  for
indefinite mass systems.    Of  course,  by taking the on-shell condition the
usual notion of proper  time  is  recovered.    In  this  case  $s$  is not a
universal parameter in contrast with  the universality implicitly involved in
the off-shell theory [see Eq.   (3)  below].    But  it  is  only  due to the
different notions of ``event'' and ``simultaneity'' considered  in each case.
[While two events $x^{\mu}$ and $x'^{\mu}$ are simultaneous  in  the standard
case when $x_0=x'_0$, in the new framework two ``events''  $(x^{\mu}, s)$ and
$(x'^{\mu}, s')$ are ``simultaneous'' when $s=s'$ (see Ref.  \cite{notar}).]

Finally in this letter, we also derive an important relation  that  allows us
to relate the Feynman parametrization to some other parametrizations, that we
have already mentioned.

\bigskip

Let us now  briefly  discuss  the  formalism  \cite{otros}  (a  more extended
development will be given elsewhere \cite{afas}).

The states of the system are determined, at a given universal scalar ``time''
$s$, by wave functions $\Psi(x,s)$  belonging  to a linear space of spinorial
functions, defined on the space-time manifold.  This space is endowed with an
indefinite bilinear Hermitian form \cite{pe,pe2},

\begin{equation}
\langle \Phi |\Psi \rangle \equiv \int \overline{\Phi }\Psi d^4x,\label{pe}
\end{equation}
where $\overline \Phi =\Phi^{\dag}\gamma ^0 $ is the usual Dirac adjoint.  An
operator $A$ is self-adjoint according to  this  ``scalar  product''  if  $A=
\overline  A$,  where  $\overline  A$  satisfies,  by   definition,  $\langle
\phi|A|\psi\rangle  =  \langle\psi|\overline  A|\phi\rangle  ^*,  \ \ \forall
\phi,  \psi$,  e.g.,  for a spinorial operator $A$ we  have  $\overline  A  =
\gamma^0  A^{\dag}  \gamma^0$,  where  $\dag$  stands  for  the transpose and
conjugate matrix.

The ``proper time'' dynamics in the Schr\"odinger picture is provided by  the
Feynman parametrization of the Dirac equation \cite{Fey}, i.e., the evolution
of a wave function $\Psi(x,s)$ is determined by

\begin{equation}
-i\frac{d}{ds}\Psi (x, s)={\cal H}\Psi(x,s).\label{ef}
\end{equation}

The Hamiltonian ${\cal  H}$  ($\gamma^{\mu}\pi_{\mu}$  for  minimal coupling)
plays the role of  the  standard  Hamiltonian  in  the  usual nonrelativistic
theory.  Therefore, the evolution  operator  in  terms of the ``time'' $s$ is
$U(s)=e^{i{\cal H}s}$.

{}From (\ref{pe}) the spin variables $\gamma ^{\mu }$ and the orbital variables
$p_{\mu }$ and $x^{\mu }$, as well  as  the Hamiltonian, become self-adjoint.
As a consequence, the evolution operator is ``unitary.'' This fact guarantees
that  the  ``norm''  is  a  constant  of motion.   (Barut  and  Thacker  have
considered  the  same parametrization but they have defined a scalar  product
which does not preserve the norm.  See Ref.  \cite{otros}.)

The evolution of an operator $q$, in the Heisenberg picture, is given by

\begin{equation}
\frac{dq}{ds}=-i[{\cal H},q].\label{db}
\end{equation}

The generalized eigenfunctions of ${\cal H}$ are definite  mass  states $\phi
_m$ that satisfy the generalized eigenvalue equation, ${\cal H}\phi  _m  =  m
\phi  _m$,  having  oscillatory behavior in $s$.  They are  solutions  of  an
extended  Dirac  equation (note that $m$ could be, in principle, any  complex
number if the norm of $\phi_m$ vanishes, or any real number if  the  norm  of
$\phi_m$  is different from zero \cite{pepd}).  If we assume that the orbital
operators have  real  eigenvalues\footnote{For example, in Eq.  (\ref{pe}) is
implicit that only  the states associated with real eigenvalues contribute to
the  spectral  decomposition  of  the  identity  operator  in  terms  of  the
generalized eigenvectors of $x^{\mu}$.} (which  will  be  considered from now
on), in the free case $m$  can  take  the  continuous  real  values  when the
generalized eigenvalues $m^2$ of ${\cal H}^2 =  p^{\mu} p_{\mu}$ are positive
(tardyons)  and the continuous purely imaginary values when  the  generalized
eigenvalues $m^2$ of ${\cal H}^2 = p^{\mu} p_{\mu}$ are  negative (tachyons).
In the last case it can be easily checked that  the  generalized eigenvectors
of ${\cal H}$ corresponding to tachyons have zero norm.

We want to show  that  the theoretical framework we have given to the Feynman
parametrization  is  not  only  an    alternative  formalism  to  the  second
quantization  of the Dirac field \cite{Fey}  but  it  allows  recovering  the
standard results of relativistic mechanics as well.    In order to show this,
we shall consider the restriction of the formalism  to  the ``positive mass''
subspace  \cite{notar}.    It  means  that  the state of  the  system  $\Psi$
satisfies   $\Lambda  \Psi  =  \Psi$,  where  the  ``projector''  \cite{proj}
$\Lambda$ is given by

\begin{equation}
\Lambda \equiv \frac{1}{2} (1+\frac{{\cal H}}{\sqrt{{\cal H}^2}}),\label{p}
\end{equation}
which is  a  straightforward  extension  of  the  well-known  positive energy
projector in the standard case.  However, notice that $\Lambda$ only projects
onto the space corresponding  to  the  states  with mass values with positive
real part for tardyons.   In  the case of tachyons it projects onto the space
corresponding to mass values with positive  purely  imaginary  part.  This is
the meaning we have given to ``positive  mass.'' This projection (analogously
to  what  happens  in  the  standard  case)  removes   the  ``covariant  {\it
Zitterbewegung}''  \cite{Bun}  and  it  will allow us to obtain  a  classical
theory,    which  restricted  on-shell  will  be  the  standard  relativistic
mechanics.   (An  off-shell  classical  theory  for  the  spinless  case  was
considered  in  Ref.     \cite{Mo}.    The  classical  theories  of  spinning
``particles'' corresponding to the  general  formalism  and the projected one
will be given elsewhere \cite{cq}.)

Let us begin by noting  that  in  the positive mass subspace, the Hamiltonian
reads

\begin{equation}
\Lambda {\cal H} \Lambda =\Lambda \sqrt{{\cal H}^2}\Lambda.\label{hpro}
\end{equation}
Then, we see that the Feynman Hamiltonian  (linear in the momenta) is reduced
to the squared root form, which was originally  proposed by Johnson (see Ref.
\cite{otros})  for  playing  the  role of Hamiltonian in an  indefinite  mass
context.    Actually,  Johnson  only  considered  tardyons  in his formalism.
However, this  restriction  is  too  strong.   (Notice that tachyons are also
needed in the  Fourier representation of the free Feynman propagator.) In the
free case the Hamiltonian  adopts  an expression, $\sqrt{p^\mu p_\mu}$, which
is independent of the spin.    As  the  projection  is  invariant  under  the
``proper time'' evolution (since $[{\cal H},\Lambda]=0$ and $d\Lambda/ds=0$),
from (\ref{db}) and (\ref{hpro}) we obtain

\begin{equation}
\Lambda \frac{dq}{ds} \Lambda =-i \Lambda  [\sqrt{{\cal  H}^2},  q]  \Lambda.
\label{rbj}
\end{equation}
This equation helps us to get insight into the meaning of the parametrization
(\ref{ef}).

Let us now take the classical limit  in  order  to  show  that  the formalism
restricted  to the positive mass subspace is an  off-shell  classical  theory
\cite{Mo}, in which the evolution parameter can be even  identified  with the
proper time without additional assumptions.  Besides, from this procedure  it
can  be immediately seen that such an on-shell classical theory includes  the
standard one.

To    perform    this  limit  we  use  a  generalization  of  the  well-known
quasiclassical states  (see Appendix).  First, notice that if $A$ and $B$ are
operators and $f(B)$ is an operator function, then

\begin{equation}
[A,f(B)]= \frac{1}{2} \frac{df}{dB} [A,B] + [A,B] \frac{1}{2} \frac{df}{dB} +
{\cal O} (\hbar ^2).\label{va1}
\end{equation}
Therefore, to first order in $\hbar$, we have

\begin{equation}
\Lambda \frac{dq}{ds} \Lambda =  \Lambda  \frac{1}{2}  (\frac{1}{2\sqrt{{\cal
H}^2}} [-i{\cal H}^2, q] +  [-i{\cal  H}^2,  q] \frac{1}{2\sqrt{{\cal H}^2}})
\Lambda. \label{va2}
\end{equation}
Now, we take  mean values with positive mass quasiclassical states.  It means
that we consider states  of  the  form $\Psi_c(x^{\mu}) = u \Phi_c(x^{\mu})$,
where $\Phi_c(x^{\mu})$ is a minimum  uncertainty  Gaussian  wave packet (see
Appendix) and $u$ is a constant spinor satisfying

\begin{equation}
\gamma^\mu  \langle  \pi_\mu(0)   \rangle_c\  u  =  \sqrt{\langle  \pi^\mu(0)
\rangle_c \langle \pi_\mu(0) \rangle_c}\ u ,\label{spinor}
\end{equation}
for which can be proved that

\begin{equation}
\Lambda \Psi_c = [1 + {\cal O}(\hbar)] \Psi_c .\label{pmqs}
\end{equation}
By taking account of Eq.   (A1)  given in the Appendix, we obtain for minimal
coupling (restoring $\hbar$)

\begin{equation}
\langle    \frac{dq}{ds}   \rangle_c  =  \frac{d\langle  q  \rangle_c}{ds}  =
\frac{1}{2  i\hbar \langle \sqrt{\pi^\mu \pi_\mu} \rangle_c} \langle [\pi^\mu
\pi_\mu  -    \frac{e\hbar}{2}    \sigma^{\mu\nu}   F_{\mu\nu},  q]\rangle_c.
\label{nrfb}
\end{equation}
Notice that $\vert\pi^\mu  \pi_\mu \vert \gg \vert (e\hbar/2) \sigma^{\mu\nu}
F_{\mu\nu} \vert$, so we  have neglected the spin term in the square root and
consistently retained it in the  commutator.   From Eq.  (\ref{nrfb}), we can
obtain the equation of motion on-shell  of  the classical variable $\langle q
\rangle_c$ as given by the standard relativistic mechanics, e.g., the Lorentz
force law and the Bargmann-Michel-Telegdi equations \cite{cq}.

For the sake  of  simplicity we will only consider the free case from now on.
In this case, Eq.  (\ref{nrfb}) reads

\begin{equation}
\frac{d  \langle  q  \rangle_c}{ds}=  \frac{1}{2  i \hbar \langle \sqrt{p^\mu
p_\mu} \rangle_c} \langle [ p^\mu p_\mu , q] \rangle_c.\label{mano}
\end{equation}
This equation is also a very important relation that establishes a connection
among    different  parametrizations    \cite{ptd}.        In    fact,    the
St\"uckelberg-Schwinger parametrization leads in  the  classical  limit  to a
Heisenberg equation of motion without  the  factor  $1/(2 \langle \sqrt{p^\mu
p_\mu} \rangle_c)$, which now appears in a direct way.  On the other hand, RD
and FSF have also considered a parametrization  which resembles that given in
Eq.  (\ref{mano}) using a second order Hamiltonian  [see  Eq.  (\ref{faho})].
As we have mentioned (see footnote 1), in the  classical  limit  it  would be
possible,  in  principle,  to  fix the initial conditions such that  $\langle
p^\mu  p_\mu  \rangle_c$ be equal to a desirable fixed value.   In  order  to
compare  RD and FSF  parametrizations\footnote{Strictly  speaking,  in  these
parametrizations there is no restriction  onto  the  positive  mass subspace;
then  $\langle  p^{\mu}  p_{\mu}  \rangle_c$ also  includes  $-\langle  \sqrt
{p^{\mu} p_{\mu}} \rangle_c$ values, which correspond to  the  negative  mass
subspace.} with Eq.  (\ref{mano}) on-shell, we should  identify  such a value
with $M^2$.  However, observe that the ``super-mass'' $M$  is  an ``intrinsic
parameter''  not  related {\it a priori} to any mass value  (see,  e.g.,  the
second work of Ref.  \cite{H}).  Moreover, we can also  note  that  from  the
point  of  view that RD and FSF can be derived by contracting the  de  Sitter
group \cite{Rom} (a sort of  ``super-nonrelativistic limit'') the eigenvalues
$p^{\mu} p_{\mu}$ must be smaller than  $M^2$  \cite{deS}.   This fact is not
compatible with the relation $\langle p^{\mu} p_{\mu}  \rangle_c  = M^2$.  On
the other hand, if one retains such an identification at the quantum level it
turns out to be a non-desirable feature because the Hamiltonian becomes state
dependent.    Likewise,  if  at  the  same level one identifies  $M$  with  a
particular  mass  value  (e.g.,  $m_0$,  eventually,  an eigenvalue $m$), the
indefinite mass  character  of  the  theory  is  in trouble.  This is a known
criticism made to  the  Fock  parametrization  (see  the  first paper of Ref.
\cite{otros}).

Let us now show that the parameter $s$ is reduced  to  the proper time.  From
Eq.  (\ref{rbj}) we have

\begin{equation}
\frac{d}{ds}(\Lambda       x^{\mu}\Lambda)=\Lambda\frac{dx^{\mu}}{ds}\Lambda=
\Lambda\frac{p^{\mu}}{\sqrt{p^{\alpha }p_{\alpha }}}\Lambda.\label{tia}
\end{equation}
Integrating it we obtain

\begin{equation}
\Lambda  [x^\mu  (s)-x^\mu  (0)]\Lambda =\Lambda\frac{p^\mu  }{\sqrt{p^\alpha
p_\alpha }}\Lambda s.\label{pif}
\end{equation}
If  we  now  take  mean  values  on a positive mass quasiclassical state with
$\langle p^{\mu}\rangle_c \langle p_{\mu}\rangle_c >0$  (tardyons)  up to the
first  order  in $\hbar$, and we  consider  the  rest  frame  ($\langle  \vec
p\rangle_c =0$), then

\begin{equation}
\langle [x^0(s)-x^0(0)]\rangle_c = {\rm sgn}\langle p^0
\rangle_c s,\label{tp}
\end{equation}
where ${\rm sgn} \langle p^0 \rangle_c$ is $1$ or $-1$ for an extended notion
of ``particle'' or ``antiparticle'' states, respectively.  (In  the  same way
that  ${\rm  sgn}\langle  p^0 \rangle_c$ classifies particle and antiparticle
states  in  the  mass  definite  theory, this notion is even  valid  for  the
indefinite mass  framework  \cite{notar}  [cf.  Ref.  \cite{Stu}).] Thus, the
parameter $s$ is  the  expectation  value  of  $x^0$  in  the  rest  frame in
agreement with the classical  notion  for  particle  states.    Moreover, the
St\"{u}ckelberg  interpretation  for  antiparticles,  as    particles  moving
backward in the time $x_0$, is  derived  from  (\ref{tp}).    Finally,  using
(\ref{mano}) for the four position [taking into account Eq.  (\ref{tia})], we
obtain

\begin{equation}
\frac{d}{ds}  \langle x^\mu  \rangle_c  \frac{d}{ds}\langle  x_\mu  \rangle_c
=\frac{\langle p^\mu \rangle_c \langle  p_\mu\rangle_c  }{\langle \sqrt{p^\mu
p_\mu} \rangle_c^2} = 1 \label{rtpc}
\end{equation}
(for  both, ``particle'' and ``antiparticle'' corresponding to the  off-shell
theory),  since $\langle \sqrt{p^{\mu} p_{\mu}} \rangle_c ^2 = \langle  p^\mu
\rangle_c \langle p_\mu \rangle_c$ by the factorization property.

Therefore,  {\it  ds  is reduced to the arc element of  the  indefinite  mass
system that follows the world line} $\langle x^{\mu} (s) \rangle_c $.

\vskip 1.5cm

We would like  to  thank  A.    K\'alnay and S.  Sonego for many illuminating
discussions.  We are  grateful to M.  Gadella for technical remarks.  We also
acknowledge Universidad de Buenos Aires for our research fellowships.

\appendix
\section*{}

We define  as quasiclassical state $\Psi_c(x^{\mu})$ any state that satisfies
the factorization property,

\begin{equation}
\displaystyle\lim_{\hbar\to 0}(\langle  AB\rangle_c-\langle A\rangle_c\langle
B\rangle_c)=0,\label{a1}
\end{equation}
for any orbital  (no spin) operators $A$ and $B$, where $\langle A\rangle_c =
\langle \Psi_c|A| \Psi_c\rangle / \langle \Psi_c| \Psi_c\rangle$.

Let    us   note  that    if    we    consider    states    of    the    form
$\Psi_c(x^{\mu})=u\Phi_c(x^{\mu})$ where $u$ is a constant spinor, then

\begin{equation}
\langle\Psi_c|\Psi_c\rangle                                =\overline{u}u\int
\Phi_c^*(x^{\mu})\Phi_c(x^{\mu})d^4x.\label{a2}
\end{equation}
Therefore, for any orbital operator $A$ we have that

\begin{equation}
\langle  A\rangle_c  =\int    \Phi_c^*(x^{\mu})  A(x^{\mu},  i\partial_{\mu})
\Phi_c(x^{\mu})d^4x,\label{a3}
\end{equation}
which is nothing else  than  the  spin  $0$  expression  of  the  mean  value
corresponding to the five dimensional  Galilean  invariant parametrization of
the Klein-Gordon equation \cite{Rom} [we have  taken  the normalization $\int
\Phi_c^*(x^{\mu}) \Phi_c(x^{\mu}) d^4x = 1$].  Then,  the problem of defining
a  state  satisfying  condition  (\ref{a1})  is independent of  the  ``norm''
$\langle\Psi_c|\Psi_c\rangle=\overline{u}u$, i.e., the factorization property
does not depend on the  indefiniteness  of  the metric, since the problem has
been essentially reduced to the spin  $0$ case mentioned above where the norm
is positive definite.  In this case,  a class of states satisfying (\ref{a1})
was previously introduced by Cooke (Ref.  \cite{DV}),


\begin{eqnarray}
\Phi_c(x^{\mu})=\frac{1}{2\pi  (\Delta_c x^0\Delta_c x^1\Delta_c  x^2\Delta_c
x^3)^{1/2}}\cr
\cdot \exp\big\{-\frac{i}{\hbar}\langle              p_{\mu}(0)\rangle_c
x^{\mu}-\sum_{\mu} \big[\frac{x^{\mu}-\langle  x^{\mu}(0)\rangle_c}{2\Delta_c
x^{\mu}}\big]^2\big\},\label{co}
\end{eqnarray}
which  correspond  to  minimum  uncertainty  Gaussian wave packets ($\Delta_c
x^{(\mu)}\Delta_c  p^{(\mu)}=\hbar/2$, $\mu=0,...,3$), centered  in  $\langle
x^{\mu}(0) \rangle_c$ and $\langle p_{\mu}(0) \rangle_c$.  The proof that the
wave packets (\ref{co}) satisfy (\ref{a1}) follows  the  same steps as in the
nonrelativistic case.  A lengthy, however straightforward,  demonstration  of
this property can be made by computing $\langle x_{\mu}^n p_{\nu}^m \rangle$,
considering  that  any operator can be expressed as a  power  series  of  the
canonical variables.  For a more direct argument see Ref.  \cite{Ya2}.

As a final remark, we can note that the scalar  function (\ref{co}) loses its
Gaussian    form    as   it  is  seen  from  another  system  of  coordinates
$\{x^{\prime\mu}\}$:     $\Phi_c(x^{\mu})=\Phi_c^{\prime}(x^{\prime\mu})\not=
\Phi_c(x^{\prime\mu})$.  However,  the  property  (\ref{a1})  stands  for any
inertial  system,  since  the   Lorentz  transformation  is  unitary  in  our
formalism.  That is,

\begin{mathletters}
\label{tl}
\begin{equation}
\Psi_c^{\prime}(x^{\mu})={\cal L}\Psi_c(x^{\mu}),
\end{equation}
\begin{equation}
{\cal   L}=\exp[\frac{i}{2}  \epsilon_{\mu\nu}    (L^{\mu\nu}+    \frac{1}{2}
\sigma^{\mu\nu})],
\end{equation}
\begin{equation}
\overline{\cal L}{\cal L}={\cal L}\overline{\cal L}=I.
\end{equation}
\end{mathletters}

In fact, it is easy  to  check  that  Eq.    (\ref{a1}) is also valid for two
operators        $A^{\prime}=\overline{\cal        L}A{\cal        L}$    and
$B^{\prime}=\overline{\cal L}B{\cal L}$,

\begin{equation}
\displaystyle  \lim_{\hbar \to 0} (\langle A^{\prime}  B^{\prime}  \rangle_c-
\langle A^{\prime} \rangle_c \langle B^{\prime} \rangle_c) =0, \label{va3}
\end{equation}
and then, from Eqs. (\ref{tl}) we finally have

\begin{equation}
\displaystyle\lim_{\hbar\to    0}(\langle      AB\rangle_{c^{\prime}}-\langle
A\rangle_{c^{\prime}}\langle
B\rangle_{c^{\prime}})=0, \ \ \ \forall A, B.\label{a5}
\end{equation}

\end{document}